\documentclass[12pt]{article}
\usepackage{epsfig,amssymb,amsmath,multirow}

\usepackage{graphicx,epsfig}

\usepackage{dcolumn}
\usepackage{bm}

\catcode`\@=11
\DeclareMathAlphabet{\mathpzc}{OT1}{pzc}{m}{it}

\font\cmss=cmss12 
\def\1{\hbox{{1}\kern-.25em\hbox{l}}}
\def\bfZ{\relax{\hbox{\cmss Z\kern-.4em Z}}}

\def \be  {\begin{equation}}
\def \ee  {\end{equation}}
\def \ba  {\begin{eqnarray}}
\def \ea  {\end{eqnarray}}
\def \baa {\begin{eqnarray*}}
\def \eaa {\end{eqnarray*}}
\def \bb  {\begin {thebibliography} }
\def \eb  {\end{thebibliography}}
\def \lab #1 {\label{#1}}

\def \matrix #1 {\left(\begin{array}{cc} #1 \end{array}\right)}

\newcommand{\as}{\ifmmode\alpha_{\rm s}\else{$\alpha_{\rm s}$}\fi}
\newcommand{\asbar}{\ifmmode\bar{\alpha}_{\rm s}\else{$\bar{\alpha}_{\rm s}$}\fi}

\newcommand{\ft}[2]{{\textstyle\frac{#1}{#2}}}

\newcommand{\Z}{{\mathbb Z}}

\font\cmss=cmss12 
\def\inbar{\,\vrule height1.5ex width.4pt depth0pt}
\def\IC{\relax\hbox{$\inbar\kern-.3em{\rm C}$}}
\def\IZ{\relax{\hbox{\cmss Z\kern-.4em Z}}}
\def\IR{{\hbox{{\rm I}\kern-.2em\hbox{\rm R}}}}

\def\IP{{\hbox{{\rm I}\kern-.2em\hbox{\rm P}}}}
\def\II{\hbox{{1}\kern-.25em\hbox{l}}}

\newbox\lett\newdimen\lheight\newdimen\lwidth
\def\ontop#1#2{\setbox\lett=\hbox{#2}\lheight\ht\lett
\multiply\lheight by 12 \divide\lheight by 10\relax%
\lwidth\wd\lett \multiply\lwidth by 8 \divide\lwidth by 10\relax #2\kern-
\lwidth%
\raise\lheight\hbox{{$\scriptstyle #1$}}\kern.1ex}

\def\XXint#1#2#3{{\setbox0=\hbox{$#1{#2#3}{\int}$}
     \vcenter{\hbox{$#2#3$}}\kern-.5\wd0}}

\begin{document}

\begin{titlepage}

\thispagestyle{empty}

\vspace*{1cm}

\centerline{\large \bf All Fermion Masses and Mixings in an Intersecting D-brane World}

\vspace{1cm}

\centerline{\sc Van E. Mayes}

\vspace{10mm}

\centerline{\it Department of Physical and Applied Sciences,}
\centerline{\it University of Houston-Clear Lake}
\centerline{\it Houston, TX 77058, USA}

\vspace{2cm}

\centerline{\bf Abstract}

\vspace{5mm}

It is shown that 
neutrino mixing angles which are consistent with current experimental observations
may be naturally obtained 
in a Pati-Salam model constructed from intersecting D6 branes 
on a $T^6/(\Z_2 \times \Z_2)$ orientifold. 
The Dirac mass matrices in the model are naturally the same as those
which are obtained by imposing a $\mathbf{\Delta(27)}$ flavor symmetry, 
which allows for near-tribimaximal mixing in the neutrino sector. 
In addition, it is possible to obtain the correct mass matrices for quarks and charged leptons, 
as well as nearly the correct CKM matrix. An RGE
analysis of the neutrino mass parameters, including the seesaw mechanism assuming a specific form
for the right-handed neutrino mass matrix is performed, and it is found that the neutrino mixing angles at the
electroweak scale are 
$\theta_{12}=35.0^{\circ}$, $\theta_{23}=47.1^{\circ}$, and $\theta_{13}=8.27^{\circ}$.
In addition,the neutrino mass-squared differences are found to be $\Delta m^2_{32} = 0.00252$~eV$^2$ and 
$\Delta m^2_{21} = 0.0000739$~eV$^2$ with $m_1=0.0146$~eV, $m_2=0.0170$~eV,
and $m_3=0.0530$~eV.  These results depend slightly upon the scale at which the RGE
running goes from being that of the MSSM to that of the SM, which we interpret to be
the lightest stop mass.  The best agreement with experimental data is for
$\tilde{m}_{t_1} \approx 4.28$~TeV. This suggest that the superpartners which produce 
the strongest signal in a hadron colllider are just out of reach at the LHC. 
\end{titlepage}

\setcounter{footnote} 0

\newpage

\pagestyle{plain}
\setcounter{page} 1

\section{Introduction}

The SM exhibits an intricate pattern of mass hierarchies and mixings
between the different generations of fermions.
The pattern of neutrino mixings is one of the most interesting aspects 
of neutrino physics today. In contrast to the small quark mixing angles,
the mixing angles between neutinos appear to be quite large. 
The observation of neutrino oscillations suggests that there are small 
mass differences between the different neutrino mass states. 
At present, the masses and mixing angles for both quarks and leptons
remains completely unexplained, as well as the question of why 
they are so different from one another.

In recent years, precision measurements of the 
neutrino mixing angles as well as the squares of the mass differences
between neutrinos have been made by several experiments. 
The best estimate of the difference in the squares of the masses 
of mass eigenstates 1 and 2 was published by KamLAND in 2005:
$\Delta m^2_{21}= 0.0000739^{+0.21}_{-0.20}$~eV$^2$~\cite{Araki:2004mb,Esteban:2018azc,Esteban:2016qun,NuFit}.
In addition, the MINOS experiment measured oscillations from an intense 
muon neutrino beam, determining the difference in the squares of the masses 
between neutrino mass eigenstates 2 and 3. Current measurements indicate 
$\Delta m^2_{32}= 0.0027$~eV$^2$~\cite{Esteban:2018azc,Esteban:2016qun,NuFit},
consistent with previous results from Super-Kamiokande~\cite{Fukuda:1998fd}.
In addition, recent snalysis of cosmological results constrains the sum of the three neutrino 
masses to be $\lesssim 0.12$~eV~\cite{Vagnozzi:2017ovm}, while additional 
analysis of combined data sets results in $0.15$~eV~\cite{Giusarma:2016phn} 
and $0.19$~eV~\cite{Giusarma:2018jei} for the upper limit. Older analyses set the upper limit
slightly higher at $0.3$~eV~\cite{Thomas:2009ae,Ade:2013zuv,Battye:2013xqa}.

The lepton mixing matrix or PMNS matrix may be parameterized 
as
\begin{equation}
U_{PMNS}   = \left(\begin{array}{ccc}
c_{12}c_{13}& s_{12}c_{13} & s_{13}e^{-i\delta_{CP}}\\
-s_{12}c_{23}-c_{12}s_{23}s_{13}e^{i\delta_{CP}} & c_{12}c_{23}-s_{12}s_{23}s_{13}e^{i\delta_{CP}} &  s_{23}c_{13} \\
s_{12}s_{23}-c_{12}c_{23}s_{13}e^{i\delta_{CP}} & -c_{12}s_{23}-s_{12}c_{23}s_{13}e^{i\delta_{CP}} &  c_{23}c_{13}\end{array} \right),
\label{TBMMassMatrix}
\end{equation}
where $s_{ij}$ and $c_{ij}$ denote $\mbox{sin}~\theta_{ij}$ and $\mbox{cos}~\theta_{ij}$ respectively, 
while ${\delta_{CP}}$ is a $CP$-violating phase.

The current best-fit values for the mixing angles from direct and indirect experiments are,
using normal ordering~\cite{Esteban:2018azc,Esteban:2016qun,NuFit},
\begin{eqnarray}
\theta_{12}=&33.82^{\circ+0.78^{\circ}}_{~-0.76^{\circ}}, \\  \nonumber
\theta_{23}=&49.6^{\circ+1.0^{\circ}}_{~-1.2^{\circ}}\\  \nonumber
\theta_{13}=&~8.61^{\circ+0.13^{\circ}}_{~-0.13^{\circ}}\\  \nonumber
\delta_{CP}=&215^{\circ+40^{\circ}}_{~-29^{\circ}}\\  \nonumber
\end{eqnarray}

One of the most studied patterns of neutrino 
mixing angles is the so-called tribimaximal mixing of the form
\begin{equation}
U_{TB}   \sim  \left(\begin{array}{ccc}
\sqrt{\frac{2}{3}} & \sqrt{\frac{1}{3}} & 0\\
-\sqrt{\frac{1}{6}} &  \sqrt{\frac{1}{3}} &  -\sqrt{\frac{1}{2}} \\
- \sqrt{\frac{1}{6}} & \sqrt{\frac{1}{3}}& \sqrt{\frac{1}{2}} \end{array} \right),
\label{TBMMassMatrix}
\end{equation}
which were consistent with early data.  However, the measurment of a 
nonzero $\theta_{13}$ by Data Bay~\cite{An:2012eh} and 
Double Chooz~ \cite{Abe:2011fz}, and confirmed by RENO~\cite{Ahn:2012nd} has 
now ruled out these mixing patterns.  
Athough tribimaximal mixing is currently ruled out by experimental data,
it still may be 
viewed as a zeroth-order approximation to more general forms of 
the PMNS matrix which are also consistent with the data.  
Thus, it is still of great importance to understand the origin of
tribimaximal mixing.  

In particular, it was shown also by Ma that mass matrices leading to 
tribimaximal and near-tribimaximal mixing may be generated by 
imposing a flavour symmetry such as $\mathbf{A4}$~\cite{Ma:2005qf} 
or $\mathbf{\Delta(27)}$~\cite{Ma:2007wu}.
Specifically, a mass matrix of the form
\begin{equation}
\mathcal{M_{\nu}}   \sim  \left(\begin{array}{ccc}
 Y & X & X \\
X & y & x \\
X & x & y \end{array} \right),
\label{TBMMassMatrix}
\end{equation}
obtained by imposing an $\mathbf{A4}$ flavour symmetry  leads to tribimaximal mixing, 
while mass matrices of the form
\begin{equation}
\mathcal{M_{\nu}}   \sim  \left(\begin{array}{ccc}
f_1 v_1 & f_2 v_3 & f_2 v_2 \\
f_2 v_3 & f_1 v_2& f_2 v_1 \\
f_2 v_2 & f_2 v_1& f_1 v_3\end{array} \right),
\label{NearTBMMassMatrix}
\end{equation}
obtained by imposing an  $\mathbf{\Delta(27)}$ flavour symmetry  
may lead to near-tribimaximalal mixing as  $\mathbf{\Delta(27)}$
contains $\mathbf{A4}$ as a subgroup.  
Although generating mass matrices of this form through 
a flavour symmetry is very elegant and does provide some 
insight into the origin of the neutrino mixing angles, the
origin of the flavour symmetries required for this in a 
fundamental theory has yet to be explained.  

String theory is a leadng candidate for such a theory.
The main challenge of string phenomenology is
to exhibit at least one string vacuum that describes the physics of
our universe in detail.  Despite progress in this direction,
this has not yet been completely achieved.  In the past two decades, a
promising approach to model building has emerged involving
compactifications with D branes on orientifolds (for reviews,
see~\cite{Ura03,BluCveLanShi05, BluKorLusSti06,Mar07}).  In such
models chiral fermions---an intrinsic feature of the Standard Model
(SM)---arise from configurations with D branes located at transversal
orbifold/conifold singularities~\cite{DouMoo96} and strings stretching
between D branes intersecting at
angles~\cite{BerDouLei96,AldFraIbaRabUra00} (or, in its T-dual
picture, with magnetized D
branes~\cite{Bac95,BluGorKorLus00,AngAntDudSag00}).  A number of
non-supersymmetric intersecting D-brane models have been constructed
that strongly resemble the SM.

However, non-supersymmetric low-energy limits of string theory suffer
from internal inconsistencies of noncanceled NS-NS tadpoles, yielding
models that destabilize the hierarchy of scales~\cite{CreIbaMar02A}.
A resolution of these issues necessarily requires $\mathcal{N} = 1$
supersymmetry.  The first semirealistic models that preserve the
latter were built in Type IIA theory on a $T^6 /(\Z_2 \times
\Z_2)$ orientifold~\cite{CveShiUra01a,CveShiUra01b}.  Subsequently,
intersecting D-brane models based on SM-like,
Pati-Salam~\cite{PatSal73},
and SU(5)~\cite{GeoGla74} gauge groups
were constructed within the same framework and systematically studied
in Refs.~\cite{Leontaris:2000hh,Anastasopoulos:2010ca,CvePap06,CvePapShi02,CveLanLiLiu04,CheLiNan06} 
\footnote{see~\cite{Antoniadis:1988cm} and~\cite{Antoniadis:1990hb}
for heterotic constructions of Pati-Salam models.}.  The
statistics of 3- and 4-generation models was studied
in~\cite{Blumenhagen:2004xx, Gmeiner:2005vz}.  Phenomenologically
interesting models have also been constructed on a $T^6/(\Z_6)$
orientifold~\cite{Honecker:2004kb}.  In addition, several different
models with flipped SU(5)~\cite{Bar81} have been suggested within
intersecting D-brane scenarios~\cite{EllKanNan02,AxeFloKok03}, as well
as models with interesting discrete-group flavor
structures~\cite{Abe:2009vi}.

 Within the framework of
D-brane modeling it was demonstrated that the Yukawa matrices $Y_{abc}
\sim \exp(- A_{abc})$ arise from worldsheet areas $A_{abc}$ spanning D
branes (labeled by $a$, $b$, $c$) supporting fermions and Higgses at
their intersections~\cite{AldFraIbaRabUra00,Cremades:2003qj}.  This
pattern naturally encodes the hierarchy of Yukawa couplings.  However,
for most string constructions, Yukawa matrices are of rank one.  In
the case of D-brane models built on toroidal orientifolds, this result
can be traced to the fact that not all of the intersections at which
the SM fermions are localized occur on the same torus.  To date only
one three-generation model is known in which this problem has been
overcome~\cite{CheLiMayNan07}, and for which one can obtain mass
matrices for quarks and leptons that nearly reproduce experimental
values.  Additionally, this model exhibits automatic gauge coupling
unification at the string scale, and all extra matter can be
decoupled.  It should be commented that the rank-1 problem for
toroidal models can also potentially be solved by D-brane
instantons~\cite{Abel:2006yk, Cvetic:2009yh, Cvetic:2009ez}.  However,
the conditions for including these nonperturbative effects are very
constraining, and at present there are no concrete realizations in the
literature in which all constraints may be satisfied.

In the following, 
 shown that 
neutrino mixing angles which are consistent with current experimental observations
may be naturally obtained 
in a Pati-Salam model constructed from intersecting D6 branes 
on a $T^6/(\Z_2 \times \Z_2)$ orientifold. 
The Dirac mass matrices in the model are naturally the same as those
which are obtained by imposing a $\mathbf{\Delta(27)}$ flavor symmetry, 
which allows for near-tribimaximal mixing in the neutrino sector. 
In addition, it is possible to obtain the correct mass matrices for quarks and charged leptons, 
as well as nearly the correct CKM matrix. An RGE
analysis of the neutrino mass parameters, including the seesaw mechanism assuming a specific form
for the right-handed neutrino mass matrix is performed, and it is found that the neutrino mixing angles at the
electroweak scale are 
$\theta_{12}=35.0^{\circ}$, $\theta_{23}=47.1^{\circ}$, and $\theta_{13}=8.27^{\circ}$.
In addition,the neutrino mass-squared differences are found to be $\Delta m^2_{32} = 0.00252$~eV$^2$ and 
$\Delta m^2_{21} = 0.0000739$~eV$^2$ with $m_1=0.0146$~eV, $m_2=0.0170$~eV,
and $m_3=0.0530$~eV.  These results depend slightly upon the scale at which the RGE
running goes from being that of the MSSM to that of the SM, which we interpret to be
the lightest stop mass.  The best agreement with experimental data is for
$\tilde{m}_{t_1} \approx 4.28$~TeV. This suggest that the superpartners which produce 
the strongest signal in a hadron colllider are just out of reach at the LHC. 

\section{A Realistic MSSM} 

The configuration of D branes must obey a
number of conditions in order to be a consistent model of particle
physics.  First, the RR tadpoles vanish via the Gauss' law
cancellation condition for the sum of D-brane and cross-cap
RR-charges~\cite{BluKorLusSti06,GimPol96}:
\be
\label{RRtadpole}
\sum_{\alpha \in {\rm stacks}} N_\alpha (\pi_\alpha + \pi_{\alpha^*})
- 4 \pi_{\rm O6} = 0 \, ,
\ee
written in terms of the three-cycles $\pi_\alpha = (n^\alpha_1,
l^\alpha_1) \times (n^\alpha_2, l^\alpha_2) \times (n^\alpha_3,
2^{-\beta} l^\alpha_3)$ that wrap $(n^\alpha_j, m^\alpha_j)$ times the
fundamental cycles $([a_j], [b_j])$ of the factorizable six-torus $T^6
= \prod_{j=1}^3 T^2_{(j)}$.  Here, the first two two-tori are
rectangular: $l_j^\alpha = m_j^\alpha$ ($j = 1,2$), while the third
two-torus can be rectangular ($\beta \! = \! 0$), or tilted such that
$l_3^\alpha = 2 m_j^\alpha + n_j^\alpha$ and $\beta = 1$.  In the
T-dual picture the tilt of the third cycle $[a_3^\prime] = [a_3] +
\ft12 [b_3]$ corresponds to turning on a non-zero NS-NS two-form $B$
field. However, it becomes nondynamical under the requirement of its
invariance under the orientifold projection $\Omega
\mathcal{R}$~\cite{BluKorLusOtt01}. As a consequence, its flux can
admit only two discrete values, resulting in two discrete values for
$\beta$. Each two-torus possesses the complex structure modulus
$\chi_j = R^{(j)}_2/R^{(j)}_1$ built from its radii $R^{(j)}_1$ and
$R^{(j)}_2$.  $\mathcal{N}=1$ supersymmetry, which is favored for
reasons of underlying consistent low-energy theories of particle
physics as well as for stability of D-brane configurations, is
preserved by choosing the angles between the D-brane stacks and
orientifold planes to obey the
condition~\cite{CveShiUra01a,CveShiUra01b}
\be
\theta^\alpha_1 + \theta^\alpha_2 + \theta^\alpha_3 = 0 \ \mbox{mod} \,
2\pi \, ,
\ee
with $\theta^\alpha_j = \arctan (2^{- \beta_j} \chi_j
l^\alpha_j/n^\alpha_j)$ and $\beta_{1,2} = 0$ and $\beta_3 =
\beta$. This condition can be written in terms of wrapping numbers 
satisfying the two equations
\begin{eqnarray}
x_A \tilde{A}_a + x_B \tilde{B}_a + x_C \tilde{C}_a + x_D \tilde{D}_a = 0, \nonumber \\
A_a/x_A + B_a/x_B + C_a/x_C + D_a/x_D < 0 ,
\end{eqnarray}
where
\begin{eqnarray}
\tilde{A}_a = -l^1_a l^2_a l^3_a, \ \ \ \tilde{B}_a = l^1_a n^2_a n^3_a, \ \ \ \tilde{C}_a = n^1_a l^2_a n^3_a, \ \ \ \tilde{D}_a = n^1_a n^2_a l^3_a, \nonumber \\
 A_a = -n^1_a n^2_a n^3_a, \ \ \ B_a = n^1_a l^2_a l^3_a, \ \ \ C_a = l^1_a n^2_a l^3_a, \ \ \ D_a = l^1_a l^2_a n^3_a,
\end{eqnarray}
and $x_A$, $x_B$, $x_C$, and $x_D$ are the complex structure
parameters~\cite{CvePapShi02}, where $x_A = \lambda$, $x_B = \lambda
\cdot 2^{\beta_2 + \beta_3}/\chi_2 \chi_3$, $x_C = \lambda \cdot
2^{\beta_1 + \beta_3}/\chi_1 \chi_3$, $x_D = \lambda \cdot 2^{\beta_ +
\beta_2}/\chi_1 \chi_2$, and $\lambda$ is a positive parameter that
puts the parameters $A$, $B$, $C$, and $D$ on equal footing.
Furthermore, the consistency of the model is further ensured by the
K-theory conditions~\cite{Mar03,BluCveMarShi05}, which imply the
cancellation of the $Z_2$ charges carried by D branes in orientifold
compactifications in addition to the vanishing of the total
homological charge exhibited by Eq.\ (\ref{RRtadpole}). In the present
case, nonvanishing torsion charges are avoided by considering stacks
with an even number of D branes, {\it i.e.}, $N_\alpha \in 2
\mathbb{Z}$.

\begin{table}[t]
\caption{General spectrum for intersecting D6 branes at generic
angles, where $I_{aa'}=-2^{3-\beta}\prod_{i=1}^3(n_a^il_a^i)$, and
$I_{aO6}=2^{3-\beta}(-l_a^1l_a^2l_a^3
+l_a^1n_a^2n_a^3+n_a^1l_a^2n_a^3+n_a^1n_a^2l_a^3)$.  ${\cal M}$ is the
multiplicity, and $a_S$ and $a_A$ denote the symmetric and
antisymmetric representations of U$(N_a/2)$, respectively.}
\renewcommand{\arraystretch}{1.4}
\begin{center}
\begin{tabular}{|c|c|}
\hline {\bf Sector} & \phantom{more space inside this box}{\bf
Representation}
\phantom{more space inside this box} \\
\hline\hline
$aa$   & U$(N_a/2)$ vector multiplet  and 3 adjoint chiral multiplets  \\
\hline $ab+ba$   & $ {\cal M}(\frac{N_a}{2},
\frac{\overline{N_b}}{2})=
I_{ab}=2^{-\beta}\prod_{i=1}^3(n_a^il_b^i-n_b^il_a^i)$ \\
\hline $ab'+b'a$ & $ {\cal M}(\frac{N_a}{2},
\frac{N_b}{2})=I_{ab'}=-2^{-\beta}\prod_{i=1}^3(n_{a}^il_b^i+n_b^il_a^i)$ \\
\hline $aa'+a'a$ &  ${\cal M} (a_S)= \frac 12 (I_{aa'} - \frac 12
I_{aO6})$~;~~ ${\cal M} (a_A)=
\frac 12 (I_{aa'} + \frac 12 I_{aO6}) $ \\
\hline
\end{tabular}
\end{center}
\label{IBspectrum}
\end{table}

Imposing these constraints, we present the D6-brane configurations,
intersection numbers, and complex structure parameters of the model in
Table~\ref{MI-Numbers}, and the resulting spectrum in
Table~\ref{SpectrumB}, with formulas for calculating the multiplicity
of states in bifundamental, symmetric, and antisymmetric states shown
in Table~\ref{IBspectrum}.  Models with different numbers of
generations may be obtained for different values of the wrapping
number $n_g$ as well as the third-torus tilt parameter $\beta$.  The
observable sector of the models then has the gauge symmetry and matter
content of an $(N_g \! = \!  2^{1-\beta} n_g)$-generation SM with an
extended Higgs sector.  The extra matter in the models consists of
matter charged under the hidden-sector gauge groups, and vectorlike
matter between pairs of branes that do not intersect, as well as the
chiral adjoints associated with each stack of branes.  In addition,
one has matter in the symmetric triplet representation of SU(2)$_L$ as
well as additional singlets.  In order to have just the MSSM at low
energies, the gauge couplings must unify at some energy scale, and all
extra matter besides the MSSM states must become massive at
high-energy scales.  Furthermore, one requires just one pair of Higgs
doublets.

The resulting models have gauge symmetry $[{\rm U}(4)_C \times {\rm
U}(2)_L \times {\rm U} (2)_R]_{\rm observable} \times [ {\rm
USp}(2^{2-\beta}(4-n_g))^2 \times {\rm USp}(2^{2-\beta})^2]_{\rm
hidden}$. The hidden sector, as well as the set of complex structure parameters
required to preserve $\mathcal{N}=1$ supersymmetry, is different in each
of the models with
different numbers of generations.  In particular, in the tilted case
two of the hidden-sector gauge groups fall out in going from
three-generation to four-generation models. The non-Abelian chiral
anomalies vanish as a consequence of the RR tadpole condition
(\ref{RRtadpole}). The chiral anomalies from the three global U(1)s 
of U(4)$_C$, U(2)$_L$, and U(2)$_R$ inducing couplings of the form 
$A_\alpha \wedge F_\beta^2$, with $A$ and $F$ referring to Abelian and 
non-Abelian gauge fields, respectively, read~\cite{IbaMarRab01}
\be
\mathcal{A}^{\rm chiral}
= 
\ft12 \sum_{\alpha, \beta} N_\alpha (I_{\alpha\beta} -
I_{\alpha^\ast \beta})  A_\alpha \wedge F_\beta^2
\, ,
\ee
However, these
anomalies cancel against the couplings induced by RR fields via the
Green-Schwarz mechanism~\cite{IbaMarRab01}:
\be
\mathcal{A}^{\rm RR}
=
8 n_g A_a \wedge \left( F_c^2 - F_b^2 \right)
+
4 n_g A_b \wedge F_a^2 
-
4 n_g A_c \wedge F_a^2
\, ,
\ee
such that $\mathcal{A}^{\rm chiral} + \mathcal{A}^{\rm RR} = 0$. The
gauge fields $A_\alpha$ of these U(1)s receive masses via linear
$\sum_\ell c_\ell^\alpha B_2^\ell \wedge A_\alpha$ couplings in the
ten-dimensional action, with the massless modes given by ${\rm ker}
(c_\ell^\alpha)$.  The latter is trivial in the present model, which
means that the effective gauge symmetry of the observable sector is
${\rm SU}(4)_C \times {\rm SU} (2)_L \times {\rm SU}(2)_R$.

\begin{figure}
	\centering
		\includegraphics[width=0.75\textwidth]{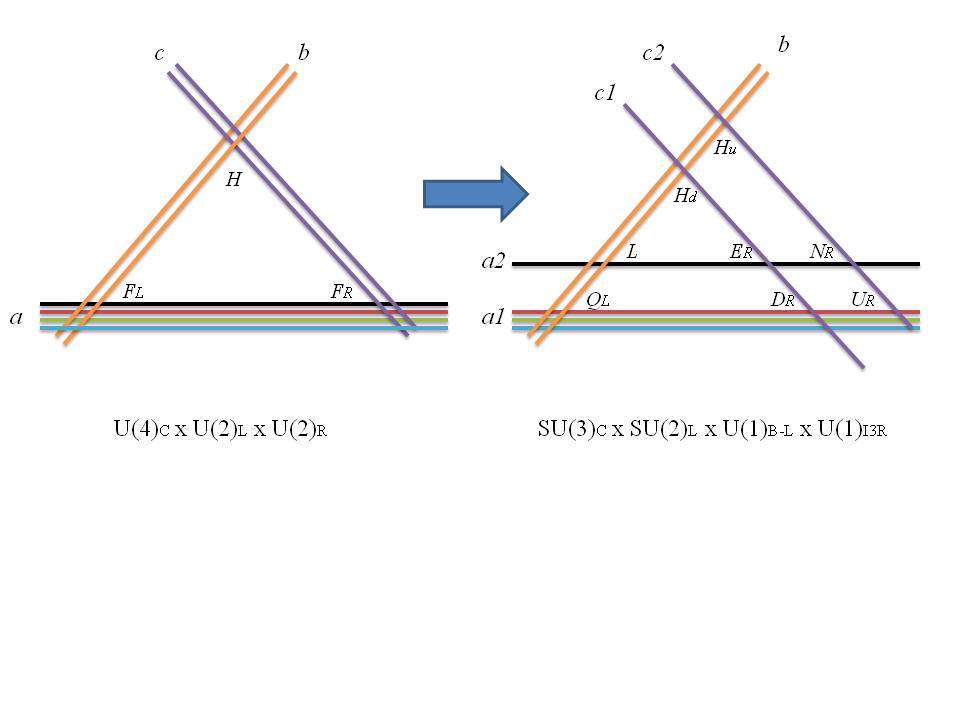}
		\caption{Breaking of the effective gauge symmetry via D-brane splitting. This process
		corresponds to assigning VEVs to adjoint scalars, which arise as open-string moduli
		associated with the positions of stacks {\it a} and {\it c} in the internal space.}
	\label{fig:Dsplitting}
\end{figure}

In order to break the gauge symmetry of the observable sector down to
the SM, we split the $a$ stack of D6 branes on the first two-torus
into stacks $a_1$ and $a_2$ with $N_{a_1}=6$ and $N_{a_2}=2$ D6
branes, and similarly split the $c$ stack of D6 branes into stacks
$c_1$ and $c_2$ such that $N_{c_1} = 2$ and $N_{c_2} = 2$ as shown in 
Fig.~\ref{fig:Dsplitting}.  The
process of brane-splitting corresponds to giving a vacuum expectation
value (VEV) to the chiral adjoint fields associated with each stack,
which are open-string moduli.  The gauge symmetry subsequently breaks
down to ${\rm SU}(3)_C \times {\rm SU}(2)_L \times {\rm U}(1)_{I_{3R}}
\times {\rm U}(1)_{B-L}$, where the U(1)$_{I_{3R}}$
and U(1)$_{B-L}$ gauge bosons remain massless.  The ${\rm
U}(1)_{I_{3R}} \times {\rm U}(1)_{B-L}$ gauge symmetry may then be
broken to ${\rm U}(1)_Y \! = \! \frac 1 2 {\rm U}(1)_{B-L} \!  + {\rm
U}(1)_{I_{3R}}$ by giving VEVs to the vectorlike particles with the
quantum numbers $({\bf 1}, {\bf 1}, 1/2, -1)$ and $({\bf 1}, {\bf 1},
-1/2, 1)$ under the ${\rm SU}(3)_C\times {\rm SU}(2)_L\times {\rm U}
(1)_{I_{3R}} \times {\rm U}(1)_{B-L}$ gauge symmetry arising from $a_2
c_1^\prime$ intersections. The full gauge symmetry of the models is
then ${\rm SU}(3)_C \times {\rm SU}(2)_L\times {\rm U}(1)_Y \times [
{\rm USp} [2^{2-\beta}(4-n_g)]^2 \times {\rm USp}(2^{2-\beta})^2]$,
with the hypercharge given by
\begin{equation}
Q_Y = \ft{1}{6} \left( Q_{a_1} - 3 Q_{a_2} - 3 Q_{c_1} + 3 Q_{c_2}
\right) \, ,
\end{equation}
where the $a$-stack charges provide $Q_{B-L}$ and the $c$-stack
charges provide $Q_{3R}$.  

\begin{table}[t]
\footnotesize
\renewcommand{\arraystretch}{1.3}
\caption{D6-brane configurations and intersection numbers for a series
of Pati-Salam models with $2^{1-\beta}n_g$ generations on a Type IIA
$T^6 / (\Z_2 \times \Z_2)$ orientifold, where the tadpole conditions
are satisfied without introducing fluxes.  The parameter $\beta$ can
be zero or one if the third torus is untilted or tilted respectively,
while the wrapping number $n_g$ may take the values $1$, $2$, $3$, or
$4$.  The complete gauge symmetry is $[{\rm U}(4)_C \times {\rm
U}(2)_L \times {\rm U}(2)_R]_{\rm observable} \times \left\{ {\rm
USp}[2^{2-\beta}(4-n_g)]^2 \times {\rm USp}
(2^{2-\beta})^2\right\}_{\rm hidden}$, and the complex structure
parameters that preserve $\mathcal{N}=1$ supersymmetry are $x_A =
x_B=n_g\cdot x_C = n_g \cdot x_D$.  The parameters $\beta^g_i$ give 
the $\beta$-functions for the hidden-sector gauge groups.}
\label{MI-Numbers}
\begin{center}
\begin{tabular}{|@{\,}c@{\,}||@{\,}c@{\,}|@{\,}c@{\,}||@{\,}c@{\,}|@{\,}c@{\,}|@{\,}c@{\,}|@{\,}c@{\,}|@{\,}c@{\,}|@{\,}c@{\,}|@{\,}c@{\,}|@{\,}c@{\,}|@{\,}c@{\,}|@{\,}c@{\,}|}
\hline
& \multicolumn{12}{c|}{${\rm U}(4)_C\times {\rm U}(2)_L\times
{\rm U}(2)_R \times {\rm USp}[2^{2-\beta}(4-n_g)]^2 \times {\rm USp}(2^{2-\beta})^2$}\\
\hline \hline  & $N$ & $(n^1,l^1)\times (n^2,l^2)\times

(n^3,l^3)$ 
& $n_{S}$& $n_{A}$ & $b$ & $b'$ & $c$ \ & $c'$& 1 \ & 2 \ & 3 \ & 4 \ \\
\hline
$a$&  8& $(0,-1)\times (1,1)\times (1,1)$ & 0 & 0  & $2^{1-\beta}n_g$ & 0 & $-2^{1-\beta}n_g$
& 0 & 1 & $-1$ & 0 & 0\\
$b$&  4& $(n_g,1)\times (1, 0)\times (1,-1)$ & $2^{1-\beta}(n_g-1)$ & $-2^{1-\beta}(n_g-1)$  & - & 0 & 0
& 0 & 0 & 1 & 0 & $-n_g$\\
$c$&  4& $(n_g,-1)\times (0,1)\times (1,-1)$ & $-2^{1-\beta}(n_g-1)$ & $2^{1-\beta}(n_g-1)$  & - & - &
- & 0 & $-1$ & 0 & $n_g$ & 0\\
\hline
1& $2^{2-\beta}(4-n_g)$& $(1,0)\times (1,0)\times (2^{\beta},0)$ & \multicolumn{10}{c|}
{$x_A = x_B = n_g \cdot x_C = n_g \cdot x_D \leftrightarrow \ \chi_1 = n_g, \ \ \chi_2 = 1, \ \ \chi_3=2^\beta$}\\
2& $2^{2-\beta}(4-n_g)$  & $(1,0)\times (0,-1)\times (0,2^{\beta})$ & \multicolumn{10}{c|}
{$\beta^g_1=-3,~\beta^g_2=-3$}\\
3& $2^{2-\beta}$ & $(0,-1)\times (1,0)\times (0,2^{\beta})$& \multicolumn{10}{c|}
{$\beta^g_3=-6+n_g$}\\    
4& $2^{2-\beta}$ & $(0,-1)\times (0,1)\times (2^{\beta},0)$& \multicolumn{10}{c|}
{$\beta^g_4=-6+n_g$}\\ 
\hline
\end{tabular}
\end{center}
\end{table}
\begin{table}
[htb] \footnotesize
\renewcommand{\arraystretch}{1.3}
\caption{The chiral and vectorlike superfields of the model, and their quantum
numbers under the gauge symmetry ${\rm U}(4)_C\times {\rm
U}(2)_L\times {\rm U}(2)_R \times {\rm USp}[2^{2-\beta}(4-n_g)]^2
\times {\rm USp} (2^{2-\beta})^2$.}
\label{SpectrumB}
\begin{center}
\begin{tabular}{|c||c||c||r|r|r||c|c|}\hline
& Multiplicity & Quantum Number & $Q_4$ & $Q_{2L}$ & $Q_{2R}$ & Field \\
\hline\hline
$ab$ & $2^{1-\beta}n_g$ & $(4,\overline{2},1,1,1,1,1)$ & 1 & $-1$ & 0  & $F_L(Q_L, L_L)$\\
$ac$ & $2^{1-\beta}n_g$ & $(\overline{4},1,2,1,1,1,1)$  & $-1$ & 0    & 1  &      $F_R(Q_R, L_R)$\\
$a1$ & 1              & $(4,1,1,\overline{N}_1,1,1,1)$ & 1 & 0 & 0  & $X_{a1}$\\
$a2$ & 1              & $(\overline{4},1,1,1,N_2,1,1)$ & $-1$ & 0 & 0   & $X_{a2}$ \\
$b2$ & 1              & $(1,2,1,1,\overline{N}_2,1,1)$ & 0 & 1 & 0    & $X_{b2}$ \\
$b4$ & $n_g$          & $(1,\overline{2},1,1,1,1,N_4)$ & 0 & $-1$ & 0    & $X_{b4}^i$ \\
$c1$ & 1              & $(1,1,\overline{2},N_1,1,1,1)$ & 0 & 0 & $-1$    & $X_{c1}$ \\
$c3$ & $n_g$          & $(1,1,2,1,1,\overline{N}_3,1)$ & 0 & 0 & 1   &  $X_{c3}^i$ \\
$b_{S}$ & $2^{1-\beta}(n_g-1)$ & $(1,3,1,1,1,1,1)$ & 0 & 2 & 0   &  $T_L^i$ \\
$b_{A}$ & $2^{1-\beta}(n_g-1)$ & $(1,\overline{1},1,1,1,1,1)$ & 0 & $-2$ & 0   & $S_L^i$ \\
$c_{S}$ & $2^{1-\beta}(n_g-1)$ & $(1,1,\overline{3},1,1,1,1)$ & 0 & 0 & $-2$   & $T_R^i$ \\
$c_{A}$ & $2^{1-\beta}(n_g-1)$ & $(1,1,1,1,1,1,1)$ & 0 & 0 & 2   & $S_R^i$ \\
\hline\hline
$ab'$   & $n_g$ & $(4,2,1,1,1,1,1)$ & 1 & 1 & 0  & \\
&        $n_g$ & $(\overline{4},\overline{2},1,1,1,1,1)$ & $-1$ & $-1$ & 0 & \\
\hline
$ac'$   & $n_g$ & $(4,1,2,1,1,1,1)$ & 1 & 0 & 1  & $\Phi_i$ \\
&        $n_g$ & $(\overline{4}, 1, \overline{2},1,1,1,1)$ & $-1$ & 0 & $-1$ & $\overline{\Phi}_i$\\
\hline
$bc$    & $2 n_g$ & $(1,2,\overline{2},1,1,1,1)$ & 0 & 1 & $-1$   & $H_u^i$, $H_d^i$\\
        & $2 n_g$ & $(1,\overline{2},2,1,1,1,1)$ & 0 & $-1$ & 1   &  \\
\hline
\end{tabular}
\end{center}
\end{table} 

The gauge coupling constant associated with a stack $\alpha$ is given
by
\begin{equation}
\label{idb:eq:gkf}
g_{{\rm D6}_\alpha}^{-2} = | \Re{\rm e} \,(f_\alpha) | 
\, , 
\end{equation}
where $f_\alpha$ is the holomorphic gauge kinetic function associated
with stack $\alpha$, given~\cite{BluKorLusSti06,CreIbaMar02A} in terms
of NS-NS fields by:
\begin{eqnarray}
f_\alpha &=& \frac{1}{4\kappa_\alpha}
\left[ n^\alpha_1\, n^\alpha_2\,n^\alpha_3 \,s 
- 
2^{-\beta} n^\alpha_1\,l^\alpha_2 \, l^\alpha_3 \, u^1 
- 
2^{-\beta} n^\alpha_2 \, l^\alpha_1 \, l^\alpha_3 \, u^2
- 
n^\alpha_3 \, l^\alpha_1 \, l^\alpha_2 \,u^3 \right] 
\, ,
\label{kingaugefun}
\end{eqnarray}
where $\kappa_\alpha = 1$ for SU($N_\alpha$) and $\kappa_\alpha = 2$
for USp($2N_\alpha$) or SO($2N_\alpha$) gauge groups.  The holomorphic
gauge kinetic function associated with SM hypercharge U(1)$_Y$ is then
given by taking a linear combination of the holomorphic kinetic gauge
functions from all of the stacks~\cite{BluLusSti03}:
\begin{equation}
f_Y = \ft{1}{6}f_{a_1} + \ft{1}{2} \left( f_{a_2} + f_{c_1} + f_{c_2}
\right).
\end{equation}
Note that in Eq.~(\ref{kingaugefun}), the four-dimensional dilaton $s$
and complex structure moduli $u^i$ refer to the supergravity basis.
These moduli must be stabilized, and gaugino condensation of the
effective Veneziano-Yankielowicz Lagrangian~\cite{VenYan82} provides
an example of such a mechanism~\cite{CveLanWan03}.  Gaugino
condensation in the hidden sectors can play an important role in
moduli stabilization, and it might provide a top-down reason why three
generations is preferred over four.

From the complex structure parameters, the complex structures $U^i$ are determined 
to be
\begin{equation}
U^1 = n_g \cdot i, \ \ \ \ \  U^2 = i, \ \ \ \ U^3 = -\beta + i.   
\end{equation}
The dilaton and complex structure moduli are then given in the supergravity basis by\footnote{See,
e.g., footnote 5 of Ref.\  \cite{BluLusSti03} for the relation between these and complex structures $U^i$.}
\begin{eqnarray}
\mbox{Re}(s) &=& \frac{1}{\left( 2^\beta n_g \right)^{1/2}} \frac{e^{-\phi_4}}{2\pi} \, , \ \ \ \ \
\mbox{Re}(u^1) \ = \left( \frac{2^\beta}{n_g} \right)^{1/2} \frac{e^{-\phi_4}}{2\pi} \, , \ \ \nonumber \\ 
\mbox{Re}(u^2) &=& \left( 2^\beta n_g \right)^{1/2} \frac{e^{-\phi_4}}{2\pi} \, , \ \ \
\mbox{Re}(u^3)  =  \frac{1}{\left( 2^\beta n_g \right)^{1/2}} \frac{e^{-\phi_4}}{2\pi} \, ,
\end{eqnarray}
where $\phi_4 = \ln g_s$ is the four-dimensional dilaton.  Inserting
these expressions into Eq.~(\ref{idb:eq:gkf}), one finds that the
gauge couplings are unified as $g^2_{s} = g^2_{w} = \frac{5}{3}g^2_Y =
g^2$ at the string scale $M_X$,
\begin{equation}
\frac{g^2 (M_X)}{4 \pi}
=
\left( \frac{2^\beta}{n_g} \right)^{1/2}e^{\phi_4}, 
\end{equation}
with the value of $\phi_4$ fixed by the value of the gauge couplings
where they unify, $g^2(M_X)$, which assumes different values for models
with different numbers of generations at $M_X =2.2\times 10^{16}$~GeV:
\be
g^2|_{N_{g} = 1}(M_X) = 0.275
\, , \
g^2|_{N_{g} = 2}(M_X) = 0.358
\, , \
g^2|_{N_{g} = 3}(M_X) = 0.511
\, , \
g^2|_{N_{g} = 4}(M_X) = 0.895
\, . 
\ee
The corresponding string scale is then given by
\begin{equation}
M_{\rm St} = \frac{g^2(M_X)}{4 \pi} \left( \frac{n_g \pi}{2^\beta}
\right)^{1/2} M_{\rm Planck},
\end{equation}
where $M_{\rm Planck}$ is the reduced Planck scale, $2.44 \times
10^{18}$~GeV.

After fixing the value of $\phi_4$, one can then determine the values
of the gauge couplings for the hidden-sector gauge groups at the
string scale:
\begin{equation}
g^2_{{\rm USp}_j} = 2^{(4-\beta/2)} \pi n_g^{(\rho_j/2)} e^{\phi_4},
\end{equation}
where $\rho_1 = \rho_2 = +1$ and $\rho_3 = \rho_4 = -1$.  Using the
beta-function parameters $\beta_j$ in Table~\ref{MI-Numbers}, the
scale at which each hidden-sector gauge group becomes confining can be
calculated:
\begin{equation}
\Lambda_j = M_X \cdot \mbox{exp} \left\{\frac{2\pi}{-\beta_j}
\left[1-\frac{2^{\beta}\pi}{g^2(M_X) n_g^{(\rho_j +1)/2}}\right]
\right\} \, .
\end{equation}

It can then be checked that the hidden-sector gauge groups have
sufficiently negative $\beta_j$ to become confining at high-energy
scales.  To have only one pair of light Higgs doublets, as is
necessary in the MSSM in order for the gauge couplings to unify, one must 
fine-tune the mixing parameters of the Higgs doublets,
specifically by fine-tuning the $\mu$ term in the superpotential. 
In particular, the $\mu$ term and right-handed neutrino masses
which may be generated via the higher-dimensional
operators~\cite{CheLiMayNan07}:
\begin{eqnarray}
W \supset &&{{y^{ijkl}_{\mu}} \over {M_{\rm St}}} S_L^i S_R^j
H_u^k H_d^l + {{y^{mnkl}_{Nij}}\over {M^3_{\rm St}}} T_R^{m}
T_R^{n} \Phi_i \Phi_j  F_R^k  F_R^l ~,~\,
\label{eqn:HiggsSup}
\end{eqnarray}
where $y^{ijkl}_{\mu}$ and $y^{mnkl}_{Nij}$ are Yukawa couplings,
and $M_{\rm St}$ is the string scale. Thus, the $\mu$ term is TeV
scale and the right-handed neutrino masses can be in the range
$10^{10-14}$ GeV for $y^{ijkl}_{\mu} \sim 1$ and $y^{mnkl}_{Nij}
\sim 10^{(-7)-(-3)}$
where $y^{ijkl}_{\mu}$ are Yukawa couplings, $M_{\rm St}$ is the
string scale, and the singlets $S_R^j$ and triplets $T_R^j$ are 
assumed to receive
string-scale VEVs, while the VEVs of the singlets $S_L^i$ are
TeV-scale.   The exact linear combinations that give the two light
Higgs eigenstates are correlated with the pattern of Higgs VEVs
necessary to obtain Yukawa matrices for the quarks and leptons,
\begin{equation}
H_{u,d} = \sum_i \frac{v^i_{u,d}}{\sqrt{\sum(v^i_{u,d})^2}},
\end{equation}
where $v^i_{u,d} = \left\langle H^i_{u,d} \right\rangle$.  Thus, at
low energies one obtains MSSM-like models with different numbers of
generations, with gauge-coupling unification $\sim 2.2 \times
10^{16}$~GeV, and matter charged under the hidden-sector gauge groups
becomes confined into massive bound states at high-energy scales.

As has been , quantities such as gauge and Yukawa couplings
depend on the VEVs of the closed-string moduli that parametrize the
size and shape of the compactified manifold, as well as the
open-string moduli that parametrize the positions of the D6-branes in
the internal space, which are associated with the presence of three
chiral adjoints in each stack.  These VEVs should be determined
dynamically.  While it is not our goal to solve this problem in the
present work, it should be mentioned that mechanisms do exist by which
this can be accomplished.  In particular, the closed-string moduli can
be stabilized in AdS by turning on fluxes in Type
IIA~\cite{Camara:2005dc}.  In fact, this mechanism has already been
demonstrated for the three-generation model~\cite{Chen:2006gd}.  Also,
gaugino condensation in the hidden sectors can provide another source
of closed-string moduli stabilization~\cite{CveLanWan03}.  The
open-string moduli may be frozen if the D-branes wrap rigid cycles, a
possibility that can exist on the $T^6/(\Z_2 \times \Z_2)$ orientifold
with discrete
torsion~\cite{Dudas:2005jx,Blumenhagen:2005tn,Forste:2010gw}.  An
example of a four-generation MSSM-like model constructed from
D6-branes wrapping rigid cycles is given in~\cite{CheLiNan06}.  We
emphasize the possibility of finding a dynamical reason to explain why
nature chooses a specific number of chiral generations by studying the
moduli stabilization problem for models with different numbers of
generations, such as our mini-landscape of models.

\section{Yukawa Couplings.} 
As one can see from the previous section
(note the filler brane stacks in Table~\ref{SpectrumB}), only the
models with $n_g \leq 4$ can satisfy the tadpole conditions without
introducing fluxes.  If we take this condition as a constraint, then
the only viable models from the top-down point of view have $N_g = 1,
2, 3, 4, 6$, and $8$.  Furthermore, masses may be generated via
trilinear couplings for all generations only for those models with a
tilted third torus ($\beta = 1$).  If we also take this condition as a
constraint, then the only viable models are those with $N_g = 1, 2,
3$, and $4$.  Additionally, the SU(3)$_C$ factor in the SM gauge group
is only asymptotically free for SUSY models with four generations or
less.  Thus, the maximum viable number of generations is four.

The three-generation model has previously been studied
in~\cite{CheLiMayNan07}.  As mentioned in the Introduction, this model
exhibits rank-3 Yukawa matrices and it is possible to nearly reproduce
the correct masses and mixings for the three known generations of
quarks and leptons. However, the Dirac mass matrix
for neutrinos was not considered. Furthermore, there were difficulties obtaining
the correct muon and electron masses which required additional corrections
from four-point functions~\cite{Chen:2008rx}.  

\begin{figure}
	\centering
		\includegraphics[width=0.5\textwidth]{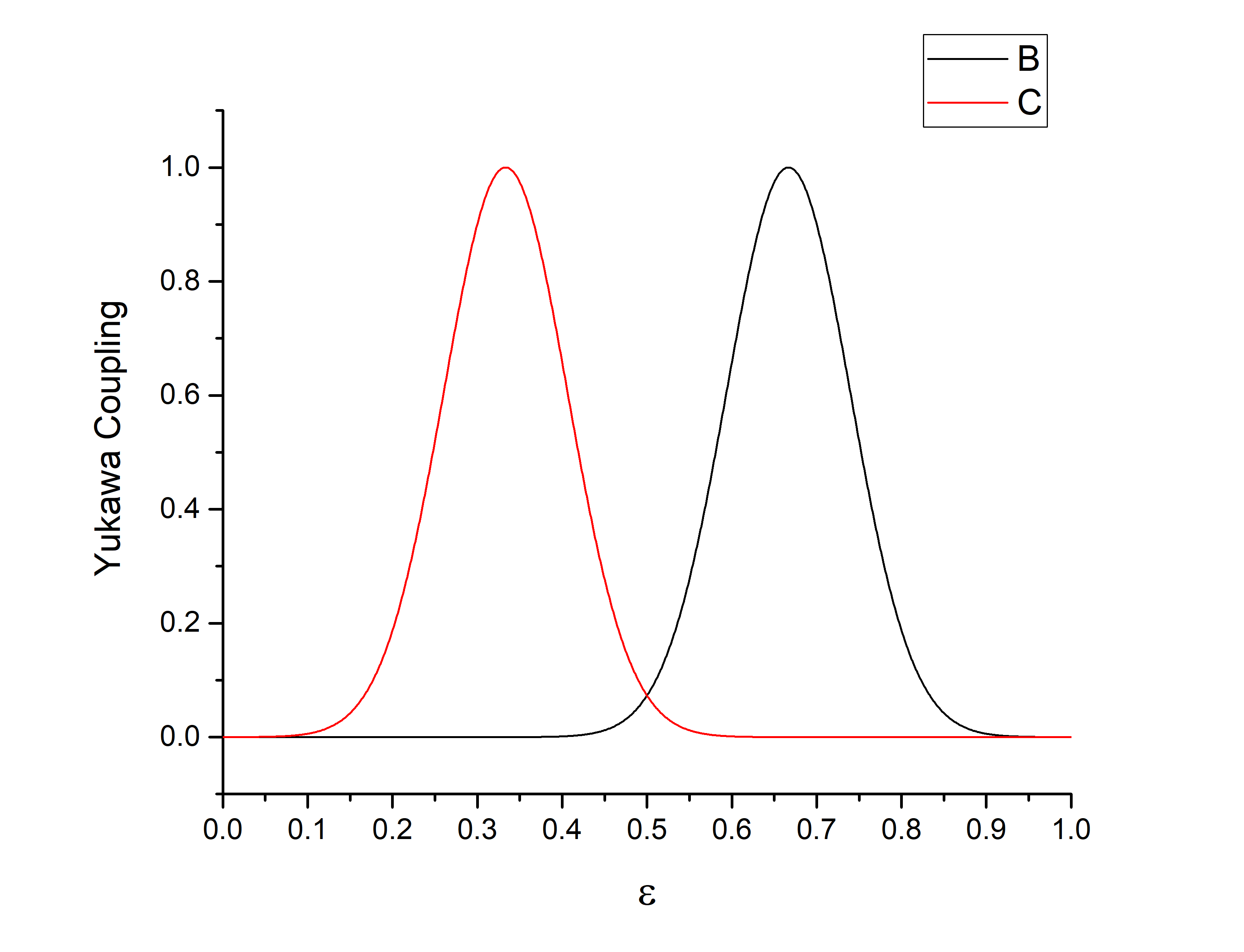}
		\caption{Yukawa couplings $ B$ and $C$ as a function of shift parameter $\epsilon$. Here it may be seen that $B=C$  for 
$\epsilon = 0$ and $\epsilon = 1/2$. }
	\label{fig:YukawaBC}
\end{figure}

\begin{figure}
	\centering
		\includegraphics[width=0.5\textwidth]{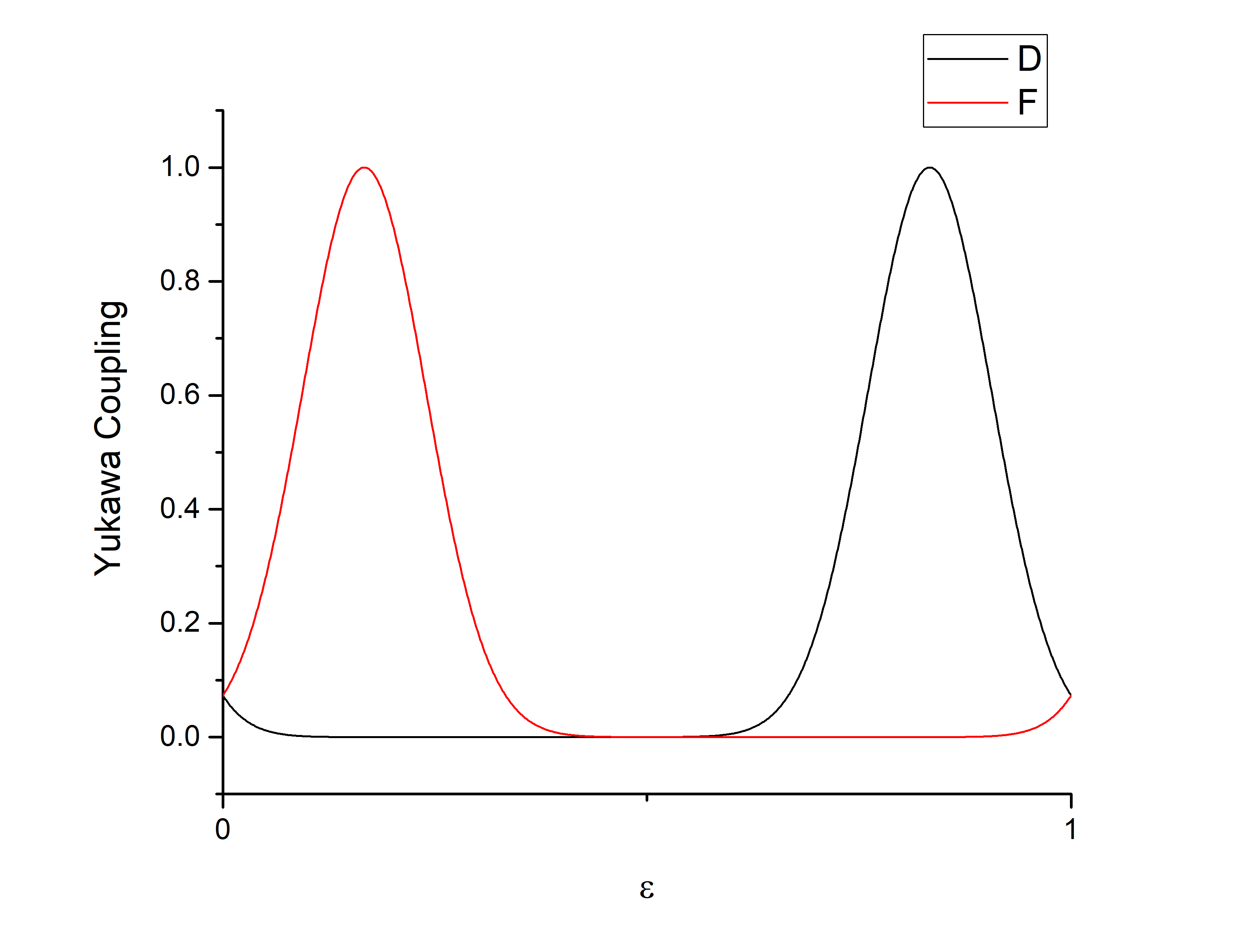}
		\caption{Yukawa couplings  amd $D$ amd $ F$  as a function of shift parameter $\epsilon$. Here it may be seen that  $D=F$ for 
$\epsilon = 0$ and $\epsilon = 1/2$. }
	\label{fig:YukawaDF}
\end{figure}

\begin{figure}
	\centering
		\includegraphics[width=0.5\textwidth]{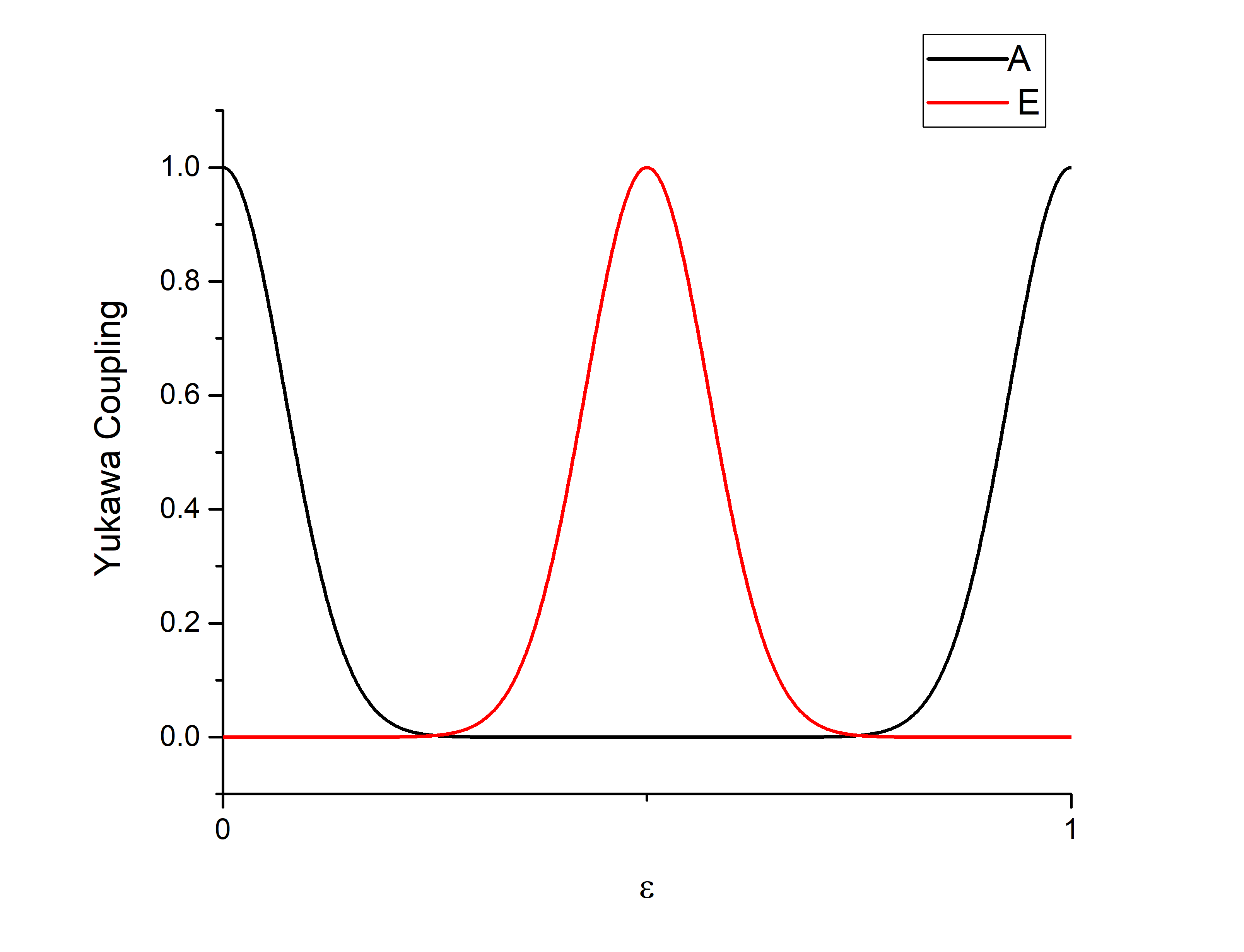}
		\caption{Yukawa couplings  amd $A$ amd $ E$  as a function of shift parameter $\epsilon$. Here it may be seen that  $A=1$ and $E=0$ for 
$\epsilon = 0$ and  $A=0$ and $E=1$ for $\epsilon = 1/2$ . }
	\label{fig:YukawaAE}
\end{figure}                

A complete form for the Yukawa couplings $y^f_{ij}$ for D6-branes
wrapping on a full compact space $T^2 \times T^2 \times T^2$ can be
expressed as~\cite{CvePap06,Cremades:2003qj}:
\begin{equation} \label{Yukawas}
Y_{\{ijk\}}=h_{qu} \sigma_{abc} \prod_{r=1}^3 \vartheta
\left[\begin{array}{c} \delta^{(r)}\\ \phi^{(r)}
\end{array} \right] (\kappa^{(r)}),
\end{equation}
where
\begin{equation}
\vartheta \left[\begin{array}{c} \delta^{(r)}\\ \phi^{(r)}
\end{array} \right] (\kappa^{(r)})=\sum_{l \in\mathbf{Z}} e^{\pi
i(\delta^{(r)}+l )^2 \kappa^{(r)}} e^{2\pi i(\delta^{(r)}+l )
\phi^{(r)}},   \label{Dtheta}
\end{equation}
with $r=1,2,3$ denoting the three two-tori.  The input parameters are
given by
\begin{eqnarray} 
\nonumber
&&\delta^{(r)} = \frac{i^{(r)}}{I_{ab}^{(r)}} +
\frac{j^{(r)}}{I_{ca}^{(r)}} + \frac{k^{(r)}}{I_{bc}^{(r)}} +
\frac{d^{(r)} ( I_{ab}^{(r)} \epsilon_c^{(r)} + I_{ca}^{(r)}
\epsilon_b^{(r)} + I_{bc}^{(r)} \epsilon_a^{(r)}
)}{I_{ab}^{(r)} I_{bc}^{(r)} I_{ca}^{(r)}} +
\frac{s^{(r)}}{d^{(r)}}, \\ \nonumber &&\phi^{(r)} =
\frac{I_{bc}^{(r)} \theta_a^{(r)} + I_{ca}^{(r)}
\theta_b^{(r)} + I_{ab}^{(r)} \theta_c^{(r)}}{d^{(r)}}, \\ 
&&\kappa^{(r)} = \frac{J^{(r)}}{\alpha'} \frac{|I_{ab}^{(r)}
I_{bc}^{(r)} I_{ca}^{(r)}|}{(d^{(r)})^2}.
\label{eqn:Yinput}
\end{eqnarray}
where the indices $i^{(r)}$, $j^{(r)}$, and $k^{(r)}$ label the
intersections on the $r^{th}$ torus, $d^{(r)} \! = gcd(I^{(r)}_{ab},
I^{(r)}_{bc}, I^{(r)}_{ca})$, and the integer $s^{(r)}$ is a function
of $i^{(r)}$, $j^{(r)}$, and $k^{(r)}$ corresponding to different ways
of counting triplets of intersections.  The shift parameters
$\epsilon_a^{(r)}$, $\epsilon_b^{(r)}$, and $\epsilon_c^{(r)}$
correspond to the relative positions of stacks $a$, $b$, and $c$,
while the parameters $\theta_a^{(r)}$, $\theta_b^{(r)}$, and
$\theta_c^{(r)}$ are Wilson lines associated with these stacks.  For
simplicity, we set the Wilson lines to zero.  
We also define the total shift parameter$\epsilon$ to be 
\begin{equation}
\epsilon=\frac{I_{ab} \epsilon_c + I_{ca} \epsilon_b + I_{bc}
\epsilon_a}{I_{ab} I_{bc} I_{ca}},
\end{equation}
so after comparing the parameters, we have
\begin{eqnarray}
&&\delta = \frac{i}{I_{ab}} + \frac{j}{I_{ca}} + \frac{k}{I_{bc}}
+ \epsilon, \\
&& \phi=0, \\
&&\kappa = \frac{J^{(r)}}{\alpha'} \frac{|I_{ab}^{(r)}
I_{bc}^{(r)} I_{ca}^{(r)}|}{(d^{(r)})^2}.
\end{eqnarray}

We focus only on the first torus, as the Yukawa couplings from the
second and third tori only produce an overall constant.  We label the
left-handed fields, right-handed fields, and Higgs fields with the
indices $i$, $j$, and $k$ respectively, which may assume the values
\begin{eqnarray}
i \in \left\{0, 1, 2 \right\}, \ \ \ \ \ j \in
\left\{0, 1, 2\right\}, \ \ \ \ \ k \in
\left\{0, 1, 2, 3, 4, 5\right\}.
\end{eqnarray}
A trilinear Yukawa coupling occurs for a given set of indices that
satisfy the selection rule
\begin{equation}
i + j + k = 0 \ \mbox{mod} \ 3. 
\label{selrule} 
\end{equation}
By choosing a different linear function for $s^{(1)}$, some independent
modes with non-zero eigenvalues are possible. Specifically, we 
will consider the case 
$s^{(1)}= -i$. This results in Yukawa matrices of the form
\begin{equation}
\mathcal{M}   \sim  \left(\begin{array}{ccc}
A v_1 & B v_3  &  C v_5 \\
C v_3 & A v_5 & B v_1 \\
B v_5 & C v_1 & A v_3 \end{array} \right)
 +
\left(\begin{array}{ccc}
E v_4 &  F v_6 & D v_2 \\
D v_6 &  E v_2 & F v_4 \\
F v_2&D v_4 & E v_6 \end{array} \right), 
\label{Yukawa general2}
\end{equation}	
where $v_k = \left\langle H_{k+1} \right\rangle$ and  
the Yukawa couplings 
$A$, $B$, $C$, $D$, $E$, and $F$ are given by
\begin{eqnarray}
&&A \equiv\vartheta \left[\begin{array}{c}
\epsilon^{(1)}\\ \phi^{(1)} \end{array} \right]
(\frac{6J^{(1)}}{\alpha'}),  \ \ \ \ \ \ \ \ \ \ \ \ \ \ \ 
B \equiv \vartheta \left[\begin{array}{c}
\epsilon^{(1)}+\frac{1}{3}\\  \phi^{(1)} \end{array} \right]
(\frac{6J^{(1)}}{\alpha'}),  \nonumber \\
&&C \equiv \vartheta \left[\begin{array}{c}
\epsilon^{(1)}-\frac{1}{3}\\  \phi^{(1)} \end{array} \right]
(\frac{6J^{(1)}}{\alpha'}), \ \ \ \ \ \ \ \ \ \ 
D \equiv \vartheta \left[\begin{array}{c}
\epsilon^{(1)}+\frac{1}{6}\\  \phi^{(1)} \end{array} \right]
(\frac{6J^{(1)}}{\alpha'}), \nonumber \\
&&E \equiv \vartheta \left[\begin{array}{c}
\epsilon^{(1)}+\frac{1}{2}\\  \phi^{(1)} \end{array} \right]
(\frac{6J^{(1)}}{\alpha'}), \ \ \ \ \ \ \ \ \ \ 
F \equiv \vartheta \left[\begin{array}{c}
\epsilon^{(1)}-\frac{1}{6}\\  \phi^{(1)} \end{array} \right]
(\frac{6J^{(1)}}{\alpha'}).
\end{eqnarray}  

These Yukawa
matrices are of rank 3, which means that it is possible to have three 
different mass eigenvalues as well as mixing between each of the different 
generations.  
 For certain values of the shift parameter $\epsilon$,
namely $\epsilon^{(1)} = 0~\mbox{mod}~0.5$, we find that some of the Yukawa couplings 
are equal due to the symmetry properties of the Jacobi Theta functions,
Eq.~(\ref{Dtheta}).  Specifically, we have that $B=C$ and $D=F$ at both $\epsilon=0$
and $\epsilon = 1/2$ as may be seen in Figs.~\ref{fig:YukawaBC} and \ref{fig:YukawaDF}.
Furthermore, at $\epsilon=0$ we find that $A=1$ while $E=0$, 
and that at $\epsilon = 1/2$ we have $A=1$ while $E=0$ as shown in Fig.~\ref{fig:YukawaAE}

Let us observe that when the shift parameters take the values $\epsilon^{(1)}=0$ 
and $\epsilon^{(1)}=1/2$, each of 
these mass matrices of the same form~Eq.~(\ref{NearTBMMassMatrix})
as given by Ma~\cite{Ma:2007wu}, which results from a $\mathbf{\Delta(27)}$
flavor symmetry. Ths is due to the symmetry properties of the Jacobi Theta functions 
and the selection rule Eq.~(\ref{selrule}).  In addition, $\mathbf{\Delta(27)}$
contains $\mathbf{A4}$ as a subroup so that it is also possible for a mass matrix 
of this form to give rise to near-tribimaximal mixing.  

It should be noted that this matrix
may be written as the sum of two matrices, one of which involves the 
odd-numbered Higgs VEVs and one involves the even-numbered Higgs VEVs, 
each of which may lead to near-tribimaximal mixing.  We shall find this useful
in the next section as the mass matrices for the up-type quarks and neutrinos
must involve the same set of Higgs VEVs $v^U_i$, and the mass matrices for
down-type quarks and charged leptons involves a different set of 
Higgs VEVs, $v^D_i$.  In particular, we will find that the up and down-type quarks
predominantly receive masses via the odd-numbered Higgs VEVs 
$v^{U,D}_{\mbox{odd}}$ while the neutrinos and charged leptons obtain
mass via the even-numbered Higgs VEVs $v^{U,D}_{\mbox{even}}$ if 
the shift parameter is $\epsilon^{(1)}=0$ for the quarks and 
$\epsilon^{(1)}=1/2$ for the leptons respectively. Finallly,
in order to have a consistent solution, the following constraint
must be satisfied
\begin{equation}
\epsilon_u + \epsilon_{l} = \epsilon_d + \epsilon_{\nu},
\end{equation}  
in order to have a consistent solution.

\section{Numerical Analysis.} 

The Yukawa couplings for the quarks and leptons are given by the superpotential
\begin{equation}
W_Y = Y^U_{ijk} Q_L^i U_R^j H_U^k + Y^D_{ijk} Q_L^i D_R^j H_D^k + Y^{\nu}_{ijk} L^i N^j H_U^k + Y^L_{ijk} L^i E^j H_D^k, 
\end{equation}
where the Yukawa couplings $Y_{ijk}$ are given by Eq.~(\ref{Yukawas})
and have the general form given by Eq.~(\ref{Yukawa general2}).

We may determine the desired mass matrices for quarks 
and leptons 
by running the RGE's up to the unification scale, which is taken to be the string scale in the
present context.
 For example, for tan$\beta \approx 50$
 at the 
unification scale $\mu=M_X$  the diagonal quark mass matrices 
$D_uU^u_L M_u {U^u_R}^{\dag}$ and $D_d=U^d_L M_d {U^d_R}^{\dag}$  
may be determined to be~\cite{Fusaoka:1998vc,Ross:2007az}
\begin{equation}
D_u = m_t \left(\begin{array}{ccc}
0.0000139 & 0 & 0  \\
0 & 0.00404 & 0 \\
0 & 0 & 1
\end{array} \right), \;\;
D_d = m_b \left(\begin{array}{ccc}
0.00141 & 0 & 0  \\
0 & 0.0280 & 0 \\
0 & 0 & 1
\end{array} \right),
\label{QMassEig}
\end{equation}
with 
\begin{equation}
V_{CKM} = U^d_L {U^u_L}^{\dag}= \left(\begin{array}{ccc}
0.9754 & 0.2205 & -0.0026i  \\
-0.2203e^{0.003^{\circ}i} & 0.9749 & 0.0318 \\
0.0075e^{-19^{\circ}i} & -0.0311e^{1.0^{\circ}i} & 0.9995
\end{array} \right),
\label{ckm}
\end{equation}
where $U^i$ are unitary diagonalization matrices.
Similarly, the diagonal charged lepton mass matrix is given by
\begin{equation}
D_l = U^l_L M_l {U^l_R}^{\dag} = m_{\tau} \left(\begin{array}{ccc}
0.000217 & 0 & 0  \\
0 & 0.0458 & 0 \\
0 & 0 & 1
\end{array} \right).
\label{CLMassEig}
\end{equation}
For the Dirac mass matrix for the neutrinos, we desire that it has the form
given in Eq.~(\ref{NearTBMMassMatrix}) so that its diagonalizion matrix 
will be near-tribimaximal.

In order to fit the mass matrices $M_u$, $M_d$, and $M_l$, we make specific choices for the 
set of Higgs VEVs $v^{U,D}_k$ as well as the shift parameter 
for each stack of D-branes, $\epsilon^{(1)}$.
In making choices for these parameters, it is useful to note that the Higgs VEVs with 
odd values of $k$ may dominate the mass matrix when the shift parameter
$\epsilon^{(1)} = 0$, while those with even values may dominate for 
$\epsilon^{(1)} = 1/2$. This is particularly true for large values of $\kappa$, 
and essentially results from the symmetry properties of the Jacobi Theta 
functions.  For example, when $\epsilon^{(1)} = 0$, the Yukawa coupling 
$A=1$ while $E=0$, but for $\epsilon^{(1)} = 1/2$ this is reversed,
$A=0$ while $E=1$ while the other Yukawa couplings $B,C,D,$ and $F$ 
tend to be smaller.  This suggest that the odd Higgs VEVs   $v^{U}_{k=odd}$
primarily give mass to the up-type quarks, while the even Higgs VEVs $v^{U}_{k=even}$
are responsible for the Dirac mass matrix for the neutrinos, provided we choose 
$\epsilon^{(1)}_u = 0$ and $\epsilon^{(1)}_{\nu} = 1/2$. 
Similarly,  the odd Higgs VEVs   $v^{D}_{k=odd}$
primarily give mass to the down-type quarks, while the even Higgs VEVs $v^{D}_{k=even}$
are responsible for the charged lepton mass matrix,
provided we choose 
$\epsilon^{(1)}_d = 0$ and $\epsilon^{(1)}_{l} = 1/2$. 
In order to have a consistent solution, we also require that the shift parameters for  
each stack of D-branes satisfies the constraint
\begin{equation}
\epsilon^{(1)}_u +\epsilon^{(1)}_l = \epsilon^{(1)}_d +\epsilon^{(1)}_{\nu} ,
\end{equation}
which the above choices clearly satisfy.

\subsection{Quark Masses and CKM Matrix}

Thus, let us make the choices $\kappa = 58.7$ and 
\begin{equation}
\begin{array}{l c l}
 v^1_u= 0.0000142, && v^1_d=0.0028224 \\
v^2_u= 0.00002408185, && v^2_d=0.045    \\
v^3_u= 1.0, && v^3_d=1.0     \\
v^4_u= 0.000000345,&& v^4_d=0.0010105      \\
v^5_u=0.00404, && v^5_d=0.0266   \\
v^6_u= 0.005960855, && v^6_d=1.0.
\end{array}
\end{equation}
In addition, let us set the shift parameters for the quarks to be $\epsilon^{(1)}_u = \epsilon^{(1)}_d = 0$. 
Note that the Higgs VEVs $v^{even}_u$ have been chosen so that the neutrino mass matrix will be near-tribimaximal.
We set all $CP$ phases to zero.
With these parameters, we obtain the following mass matrices for the up
and down-type quarks:
\begin{equation}
M_u = m_t \left(\begin{array}{ccc}
0.0000142& 0.00003553304 & 0.0000001436  \\
 0.00003553304 & 0.00404 & 0.000000002055 \\
0.0000001436 & 0.000000002055 & 1
\end{array} \right), \;\; 
\end{equation}
\begin{equation}
M_d = m_b \left(\begin{array}{ccc}
0.0028224 & 0.005960856 & 0.0002682385  \\
 0.005960856 & 0.0266 & 0.000006023448 \\
0.0002682385 & 0.000006023448 & 1
\end{array} \right).
\end{equation}
The eigenavlues for these matrices are exactly those given by Eq.~(\ref{QMassEig}), while
the diagonlization matrices for the up and down-type quarks are given by
\begin{equation}
U^u_L =  \left(\begin{array}{ccc}
0.999961& 0.008825& 0.0  \\
 -0.008825& 0.999961 & 0.0 \\
0.0 & 0.0& 1
\end{array} \right), \;\; 
\end{equation}
\begin{equation}
U^d_L = \left(\begin{array}{ccc}
0.973122& 0.230291 & 0.000269  \\
-0.023091 & 0.973122 & 0.000008 \\
-0.000268260 & -0.000070 & 1
\end{array} \right).
\end{equation}
Then the CKM matrix is given by
\begin{equation}
V_{CKM} = U^d_L {U^u_L}^{\dag}= \left(\begin{array}{ccc}
0.9751 & 0.2217 & 0.003 \\
-0.2217 & 0.9751 & 0.0 \\
-0.0003& -0.0001 & 1.0
\end{array} \right),
\end{equation}
which is very close to the desired CKM matrix Eq.~(\ref{ckm}), though not exact. 

\subsection{Lepton Masses and PMNS Matrix}

Let us set the shift parameters for the leptons to be $\epsilon^{(1)}_{\nu} = \epsilon^{(1)}_l = 1/2$.
Then, using the same set of Higgs VEVs as before, we obtain mass matrices for the 
neutrinos and charged leptons given by
\begin{equation}
M_{\nu} = m_t \left(\begin{array}{ccc}
0.000000345& 0.005960855 & 0.00002408185  \\
 0.005960855 & 0.00002408185 & 0.00000008464414\\
0.00002408185 &  0.00000008464414 & 0.005960855
\end{array} \right), \;\; 
\end{equation}
\begin{equation}
M_l = m_b \left(\begin{array}{ccc}
0.0010105 & 0.005960856 & 0.0001585588 \\
0.005960856 & 0.045 & 0.00001682392 \\
 0.0001585588 & 0.00001682392 & 1
\end{array} \right).
\end{equation}
The eigenavlues for the charged leton mass matrix are exactly those given by Eq.~(\ref{CLMassEig}), 
assuming that $m_{\tau}=m_b$. Thus, we find that it is possible to accomadate the correct masses
for quarks and charged leptons, as well as nearly the correct CKM matrix for quarks.  

\begin{figure}
	\centering
		\includegraphics[width=0.5\textwidth]{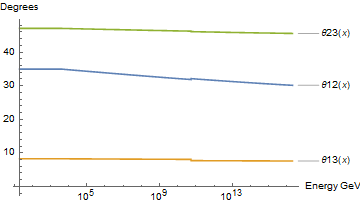}
		\caption{Neutrino mixing angles as a function of energy scale. }
	\label{fig:MixingAngles}
\end{figure}

\begin{figure}
	\centering
		\includegraphics[width=0.5\textwidth]{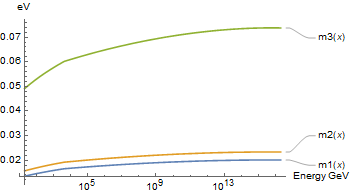}
		\caption{Neutrino masses as a function of energy scale.Note that we have taken the supersymmetry decoupling scale to be 
$4.28$~TeV in order to obtain the best agreement with data.  }
	\label{fig:NeutrinoMasses}
\end{figure}

\begin{figure}
	\centering
		\includegraphics[width=0.5\textwidth]{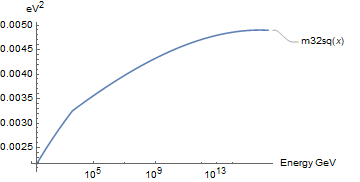}
		\caption{Mass difference $m^2_{32}$ as a function of energy scale.Note that we have taken the supersymmetry decoupling scale to be 
$4.28$~TeV in order to obtain the best agreement with data. }
	\label{fig:m32sq}
\end{figure}

\begin{figure}
	\centering
		\includegraphics[width=0.5\textwidth]{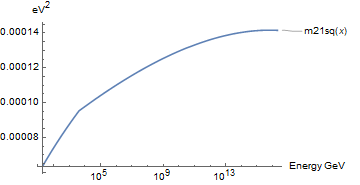}
		\caption{Mass difference $m^2_{32}$ as a function of energy scale.Note that we have taken the supersymmetry decoupling scale to be 
$4.28$~TeV in order to obtain the best agreement with data. }
	\label{fig:m21sq}
\end{figure}

The diagonlization matrices for the Dirac neutrinos and charged leptons are given by
\begin{equation}
U^{\nu}_L =  \left(\begin{array}{ccc}
0.707811& 0.578043& -0.406042 \\
 -0.706400& 0.578034 & -0.408504 \\
-0.001426 & 0.575972& 0.817468
\end{array} \right), \;\; 
\end{equation}
\begin{equation}
U^{l}_L = \left(\begin{array}{ccc}
0.991257& 0.131942 & 0.000159  \\
-0.131942 &0.9912157 & 0.000019 \\
-0.000155 & -0.000039 & 1
\end{array} \right).
\end{equation}
Here, we can see that the Dirac neutrino mass matrix is near-tribimaximal
while the charged lepton mass matrix is near-diagonal.  Of course, to explain
the observed tiny neutrino masses, usually the Dirac neutrino mass matrix
is input into a seesaw mechanism to produce a resulting Majorana mass
matrix.  In order for the canonical seesaw mechanism to work, there must
exist a mass term for right-handed neutrinos in the superpotential.  As discussed 
previously such a term is present in the model, {\it viz} Eq.~(\ref{eqn:HiggsSup}).
However, this term is of very high order, and it is not currently possible to calculate. 
In principle, such couplings may be induced by D-brane 
instantons~\cite{Abel:2006yk, Cvetic:2009yh, Cvetic:2009ez}. However, this is not
possible in the present model since the right-handed neutrino fields are charged under
a gauged $U(1)_{B-L}$.

Iin order to calculate Majorana masses for the neutrinos, we need to make some
assumptions regarding the right-handed neutrino masses.  Let us observe that
for the near tribimaximal mixing to be preserved after the seesaw mechanism is applied, 
the right-handed neutrino masses 
need to be nearly degenerate. 
Thus, let us make the choice 
\begin{equation}
M^{-1}_R = -\frac{1}{M_r}\left(\begin{array}{ccc}
0.9990& 0 & 0 \\
0& 0.99939 & 0 \\
0 & 0 & 1
\end{array} \right).
\end{equation}
where $M_r \sim 10^{10}$~GeV.
for the inverse right-handed neutrino mass matrix.
Then, applying the canonical seesaw mechanism, 
\begin{equation}
M^M_{\nu} = - M_{\nu}~M^{-1}_R~M^T_{\nu},
\end{equation}
we find that the Majorana neutrino mass matrix
is given by 
\begin{equation}
M^{\nu}_M = \left(\begin{array}{ccc}
3.5513& 0.01460&0.0144 \\
0.0146& 3.5497 & 0.0144 \\
0.0144 & 0.0144&  3.5532
\end{array} \right) \cdot10^{-3}., \;\; 
\end{equation}

Tthe diagonilization matrix for the Majorana
neutrino mass matrix remains near- tribimaximal,
\begin{equation}
U^{\nu}_M = \left(\begin{array}{ccc}
0.6086& 0.5482& 0.5737 \\
-0.7808& 0.2851 & 0.5559 \\
0.1412 & -0.7863&  -0.6015
\end{array} \right)., \;\; 
\end{equation}
In addition, 
the mass
eigvenvalues are given by
\begin{eqnarray}
m_1 = 3.5357~m_{\nu}, \ \ \ \ \ \ \ \ \  m_2 = 3.5380~m_{\nu}, \ \ \ \ \ \ \ \ \ \  m_3 = 3.5803~m_{\nu},
\end{eqnarray}
which is normal-ordered.  
If we chose  $ m_{\nu}=0.1$~eV, we find that the differences in the mass-squared values are
\begin{eqnarray}
\Delta m^2_{32}=0.003~\mbox{eV}^2, \ \ \ \ \ \ \ \ \ \ \ \ \ \ \  \Delta m^2_{21}=0.000163~\mbox{eV}^2
\end{eqnarray}
which are comparable with the results of oscillation experiments~\cite{Araki:2004mb,Fukuda:1998fd}.  
However, these results seem 
to be inconsistent with limits on the sum of neutrino masses
from cosmological data~\cite{Thomas:2009ae,Ade:2013zuv,Battye:2013xqa}. 
\begin{equation}
\sum_i \nu_i \lesssim 0.12~\mbox{eV}.
\label{upperlimit}
\end{equation}
On the the other hand, it should be remembered that these are the masses
at the string scale, rather than at  low energy where the experiments 
are performed. The RGE running of these masses could change these
results, although the change is not expected to be large.
The RGE running of the neutrino mass parameters in the
MSSM with $\mbox{tan}~\beta=50$ has been studied in~\cite{Antusch:2003kp}.

The lepton mixing matrix is given by
\begin{equation}
V_{PMNS}=  {U^{l}_L}^{\dag} U^{\nu}_M = \left(\begin{array}{ccc}
0.7063& 0.5059 & 0.4952 \\
-0.6891& 0.3590 & -0.6310 \\
0.1413& -0.7862 & -0.6016
\end{array} \right),
\end{equation} 
which may be reordered as 
\begin{equation}
V_{PMNS} =   \left(\begin{array}{ccc}
0.7862& -0.6016& 0.1413\\
0.3590&- 0.6310& -0.6891 \\
0.5059& 0.4952 & 0.7063
\end{array} \right).
\end{equation}
This result may compared to the $3\mathbf{\sigma}$ ranges on the PMNS matrix~\cite{Esteban:2018azc}:
\begin{equation}
|V|^{3\mathbf{\sigma}}_{PMNS}  = \left(\begin{array}{ccc}
0.797\rightarrow0.842& 0.518\rightarrow0.585 & 0.143\rightarrow0.156 \\
0.233\rightarrow0.495& 0.448\rightarrow0.679&0.639\rightarrow0.783 \\
0.287\rightarrow0.532& 0.486\rightarrow0.706& 0.604\rightarrow0.754
\end{array} \right).
\end{equation}

It may be observed that the absolute values of the PMNS matrix elements obtained 
in the model agree relatively well with the experimentally observed values.  
However, once again it should be kept in mind that the obtained PMNS matrix
is calculated at the string scale, while the experimentally obtained mixing angles are
obtained at low energy.  Thus, one should consider the RGE running of these 
parameters when making a true comparison.  This may be performed in the REAP
Mathematica package~\cite{Antusch:2005gp}, taking the above Yukawa matrices
for quarks and leptons as input.  
As will be discussed in detail 
in~\cite{GHM}, if we chose the right-handed neutrino mass matrix to be 
\begin{equation}
M_R = M_r\cdot\left(\begin{array}{ccc}
-5.15192& -0.279985 & 0.249735 \\
-0.279985 & -3.13943 & 1.7322 \\
  0.249735& 1.7322 & 3.15014
\end{array} \right)
\end{equation}
and start the RGE running from the conventional GUT scale $M_{GUT}=2\cdot10^{16}$~GeV
down to the electroweak scale, we obtain mixing angles which are consistent with current observations
\begin{eqnarray}
\theta_{12} = 35.0^{\circ}, \ \ \ \ \ \ \ \ \ \ 
\theta_{12} = 47.1^{\circ}, \ \ \ \ \ \ \ \ \ \ 
\theta_{13} = 8.27^{\circ}.
\end{eqnarray}
A plot of the neutrino mixing angles as a function of energy scale is 
shown in Fig.~\ref{fig:MixingAngles}. 
In addition, we find that the neutrino masses are given by 
\begin{eqnarray}
m_1 = 0.0146~\mbox{eV}, \ \ \ \ \ \ \ \ \ \ 
m_2 = 0.0170~\mbox{eV}, \ \ \ \ \ \ \ \ \ \  
m_3 = 0.0530~\mbox{eV}, \ \ \ \ \ \ \ \ \ \ 
\end{eqnarray}
with 
\begin{eqnarray}
\Delta m^2_{32} = 0.00252~\mbox{eV}^2, \ \ \ \ \ \ \ \ \ \  \ \ \ \ \ 
\Delta m^2_{21} = 0.00000739~\mbox{eV}^2.
\end{eqnarray}
These values are consistent with current experimental observations as well 
as constraints on the sum of neutrino masses from cosmological data. 
A plot of the neutrino masses as a function of energy scale is 
shown in Fig.~\ref{fig:NeutrinoMasses}. In addition, plots of $m^2_{32}$
and $m^2_{21}$ as functions of energy scale are shown in 
Fig.~\ref{fig:m32sq} and Fig.~\ref{fig:m21sq}. 
Note that we have taken the supersymmetry decoupling scale to be 
$4.28$~TeV in order to obtain the best agreement with data.  
 It would also be interesting 
to study the RGE running including light vector-like states as it has been shown
previously that such states may exist in this model in complete 
$SU(5)$ multiplets~\cite{Li:2016tqf}.
We save this for future work.

\section{Conclusion.}

We have studied the Yukawa mass matrices in a realistic MSSM constructed 
with interecting D6 branes on a $T^6/(\Z_2 \times \Z_2)$ orientifold. 
It has been shown that correct mass matrices for quarks and charged leptons 
may be obtained in the model.  Though a similar result has been demonstrated 
in previous work, before to obtain the correct 
masses for the muon and electron required additional
corrections from four-point fuctions. Here, they are obtained with only trilinear couplings.
In addition, we have obtained a CKM
quark mixing matrix which is nearly correct.  

Moreover, we have also shown that the generic Dirac mass matrices 
in the model are of the same form as those shown by Ma to lead to
near-tribimaximal mixing. We have shown that this occurs naturally in the moodel. 
Though tribimaximal mixing has been ruled out
by experiments, it still may be used as a zeroth order approximation.  
Thus, we calculate a Dirac mass matrix which is consistent with the obtained
mass matrices for quarks and leptons. In order to preserve the near-tribimaximal
mixing after the seesaw mechanism, we assumed that the right-handed neutrino 
masses are nearly degenerate. Whether or not this assumption is justified will
requie the detailed calculation of the right-handed neutrino mass matrix.  

After the seesaw mechanism, we obtained a Majorana neutrino mass matrix 
with mass eigenvalues that are such that the differences in the mass-squared
values between neutrino masses may almost be obtained. However, the sum of the 
neutrino mass eigenvalues is not consistent with the constraints from cosmological data
b roughly a factor of three.   Finally a PMNS
lepton mixing matrix is obtained with elements which are mostly consistent 
with the experimentally observed values.  It should be pointed out that this
result is dependent on small differences between the right-handed neutrino masses, 
as well as corrections from the charged lepton sector.  

We also discussed that the calculated values for the neutrino masses and PMNS matrix elements
are at the string scale,
whereas the experimentally observed values are at low-energies. We performed an RGE
analysis of the neutrino mass parameters, including the seesaw mechanism assuming a specific form
for the right-handed neutrino mass matrix, and found that the neutrino mixing angles at the
electroweak scale are consistent with experimental data. Specifically, we found that 
$\theta_{12}=35.0^{\circ}$, $\theta_{23}=47.1^{\circ}$, and $\theta_{13}=8.27^{\circ}$.
In addition, we found the neutrino mass-squared differences to be $\Delta m^2_{32} = 0.00252$~eV$^2$ and 
$\Delta m^2_{21} = 0.0000739$~eV$^2$ with $m_1=0.0146$~eV, $m_2=0.0170$~eV,
and $m_3=0.0530$~eV.  These results depend slightly upon the scale at which the RGE
running goes from being that of the MSSM to that of the SM, which we interpret to be
the lightest stop mass.  The best agreement with experimental data is for
$\tilde{m}_{t_1} \approx 4.28$~TeV. This suggest that the superpartners which produce 
the strongest signal in a hadron colllider are just out of reach at the LHC. 

Our spproach here has been to take the known  quark masses, CKM matrix, charged lepton 
masses, and the requirement for near-tribimaximal mixing for the neutrinos as input which are then
fitted in the model to fix model parameters.  This was then used to determine the Dirac mass matrix
for the neutrinos which is consistent with this fit.  We then performed an RGE analysis of the neutrino mass
parameters, which allows a prediction for the neutrino masses to be made.  
Of course, the main weakness of this approach at the moment is in the right-handed neutrino mass matrix,
which is treated as a free parameter in this analysis.  However, in principle even this may be calculated
from the model.   
We plan to focus on this  in future work.

\paragraph{Acknowledgements.}
I would like to thank my students Evan Howington, Jordan Gemmill, and Matt Teel for useful discussions during early parts of this work. 
%


\end{document}